\documentclass[fleqn,usenatbib]{mnras}

\usepackage[T1]{fontenc}

\DeclareRobustCommand{\VAN}[3]{#2}
\let\VANthebibliography\thebibliography
\def\thebibliography{\DeclareRobustCommand{\VAN}[3]{##3}\VANthebibliography}

\usepackage{graphicx}	
\usepackage{amsmath}	
\usepackage{amssymb}	
\usepackage{color}
\usepackage{float}

\def\apj {ApJ}
\def\apjl {ApJL}

\def\aj {AJ}
\def\aap {A\&A}
\def\mnras {MNRAS}

\def\nat {Nature}

\def\R200 {R_{200}}

\def\sag {\textsc {sag}}
\def\roger {\textsc {roger}}
\def\mdpl {\textsc {mdpl2}}
\def\gr {^{0.1}(g-r)}

\title[ROGER II]{Reconstructing Orbits of Galaxies in Extreme Regions (ROGER) II: reliability of projected phase-space in our understanding of galaxy populations}
\author[Coenda et al.]{\parbox[t]{\textwidth}{
Valeria Coenda,$^{1,2}$\thanks{E-mail: vcoenda@unc.edu.ar}
Mart\'in de los Rios,$^{3,4,5}$
Hern\'an Muriel,$^{1,2}$
Sof\'ia A. Cora,$^{6,7}$
H\'ector J. Mart\'inez,$^{1,2}$
Andr\'es N. Ruiz$^{1,2}$ and 
Cristian A. Vega-Mart\'inez$^{8,9}$\\}
\\
\\ 
$^{1}$Instituto de Astronom\'ia Te\'orica y Experimental (CCT C\'ordoba, CONICET, UNC), 
Laprida 854, X5000BGR, C\'ordoba, Argentina\\
$^{2}$Observatorio Astron\'omico, Universidad Nacional de C\'ordoba,
Laprida 854, X5000BGR, C\'ordoba, Argentina\\
$^{3}$ICTP South American Institute for Fundamental Research \& Instituto de F\'isica Te\'orica, Universidade Estadual \\ Paulista, 01140-070, S\~ao Paulo-SP, Brazil\\
$^{4}$ Departamento de F\'isica Te\'orica, Universidad Aut\'onoma de Madrid, 28049 Madrid, Spain\\
$^{5}$ Instituto de F\'isica Te\'orica, IFT-UAM/CSIC, C/ Nicolás Cabrera 13-15, \\ Universidad Autónoma de Madrid, Cantoblanco, Madrid 28049, Spain\\
$^{6}$Instituto de Astrof\'isica de La Plata (CCT La Plata, CONICET, UNLP), 
Observatorio Astron\'omico, Paseo del Bosque S/N, B1900FWA,\\
La Plata, Argentina\\
$^{7}$Facultad de Ciencias Astron\'omicas y Geof\'isicas, Universidad Nacional de La Plata, 
Observatorio Astron\'omico,\\
Paseo del Bosque S/N, B1900FWA, La Plata, Argentina, \\
$^{8}$Instituto de Investigaci\'on Multidisciplinar en Ciencia y Tecnolog\'ia, Universidad de La Serena, Ra\'ul Bitr\'an 1305, La Serena, Chile\\
$^{9}$Departamento de Astronom\'ia, Universidad de La Serena, Av. Juan Cisternas 1200 Norte, La Serena, Chile
}

\begin{document}
\label{firstpage}
\pagerange{\pageref{firstpage}--\pageref{lastpage}}
\maketitle

\begin{abstract}
We connect galaxy properties with their orbital classification by
analysing a sample of galaxies with stellar mass 
$M_{\star} \geq 10^{8.5}h^{-1}M_\odot$ residing in and around
massive and isolated galaxy clusters with mass $M_{200} > 10^{15}h^{-1}M_\odot$ 
at redshift $z=0$. 
The galaxy population is generated by applying the semi-analytic model of galaxy formation \sag~on the cosmological simulation MultiDark Planck 2.
We classify galaxies considering their real orbits (3D) and their projected phase-space position using the \roger~ code (2D). We define five categories:
cluster galaxies, galaxies that have recently fallen into a cluster, backsplash galaxies, infalling galaxies, and interloper galaxies.
For each class, we analyse the $\gr$ colour, the specific star formation rate (sSFR), and the stellar age, as a function of the stellar mass.
For the 3D classes, we find that cluster galaxies have the lowest sSFR, and are the reddest and
the oldest, as expected from environmental effects. 
Backsplash galaxies have properties intermediate between the cluster and recent infaller galaxies.
For each 2D class, we find an important contamination 
by other classes. We find it necessary to separate 
the galaxy populations in red and blue to perform a more realistic analysis of the 
2D data. For the red population, the 2D results are in good agreement 
with the 3D predictions. Nevertheless, when the blue population is considered, the 2D analysis only provides reliable results for recent infallers, infalling galaxies and interloper galaxies.
\end{abstract}


\section{Introduction}\label{sec:intro}

Clusters of galaxies play a key role in galaxy evolution. In the last decades there have been several studies proving strong correlations between galaxy properties and their environment (see, for instance, \citealt{ Blanton05, martinez06, martinez08}). Properties such as colour (e.g. \citealt{ Balogh:1999, Baldry:2006, Coenda:2018, Venhola:2019}), luminosity (e.g. \citealt{ Adami:1998, Girardi:2003,  Coenda:2006}), star formation rate  (e.g. \citealt{Muzzin:2012, Darvish:2016, Coenda:2019}) and morphology (e.g. \citealt{dressler80, Bamford:2009, Skibba:2009, Kawinwanichakij:2017}) correlate with the galaxy position within the cluster.

Multiple physical mechanisms affect galaxies as they move through the deep potential 
well of  a galaxy cluster.
They can experience tidal stripping
(e.g. \citealt{Zwicky:1951}; \citealt{Gnedin:2003a}; \citealt{Villalobos:2014}), which, in turn, 
can induce a central star formation burst \citep{Byrd:2001} and bar instabilities 
\citep{Lokas:2016}, remove stars from the galaxy (e.g. \citealt{Ramos:2018}), and truncate dark 
matter haloes (e.g. \citealt{Gao:2004}; 
\citealt{Limousin:2009}). In intermediate-density regions, gravitational phenomena such as 
galaxy-galaxy interactions and mergers can produce morphological transformations 
(\citealt{Moore:1996,Moore:1998,Gnedin:2003b,Smith:2015}). Additionally, as galaxies move at 
high speed through the hot ionised gas of the intracluster medium, they can have a significant
fraction of the cold gas removed through
ram pressure stripping 
(e.g. \citealt{GG:1972,Abadi:1999,Book:2010,Steinhauser:2016}). On the other hand, the warm gas
in the galaxy's halo can also be removed, a process
called starvation or strangulation (e.g. 
\citealt{Larson:1980,Balogh:2000,McCarthy:2008,Bekki:2009, Peng10, Bahe:2013, Vijayaraghavan:2015}).
The removal of these gaseous components ends up with red quenched galaxies.

In the hierarchical clustering scenario, clusters of galaxies are continuously accreting galaxies. 
In this process of falling, galaxies could undergo different physical processes that 
affect their star formation even before they reach the cluster, the so-called 
pre-processing (e.g. \citealt{Balogh:1999,Mihos:2004,Fujita:2004,McGee:2009,
deLucia:2012,Jaffe:2012,Wetzel:2013, Hou:2014, Pallero:2019}). 
Most galaxies that fall into clusters come along filaments and experience different 
environmental effects than those falling from other directions (e.g. \citealt{Martinez16,Salerno2019}). Moreover, 
galaxies that are members of a falling group will have different histories than 
those falling in isolation  (e.g. \citealt{McGee:2009, deLucia:2012, Wetzel:2013,
Hou:2014}). Thus, from where and with whom a galaxy falls into a 
cluster matters.

In recent 
years, several authors began to study the outskirts of galaxy clusters (e.g. \citealt{Mamon:2004, 
Gill:2005,Rines:2005, Aguerri:2010, Mahajan:2011, Muriel:2014, Salerno:2020, Benavides:2021}), finding not only 
infalling star-forming galaxies but also backsplash galaxies, i.e., galaxies whose orbits have taken them 
close to the cluster centre and are now beyond the virial radius. Having experienced different environmental effects during their evolution, it is expected that
backsplash galaxies and galaxies infalling for the first time 
have different properties. Both should also present
differences with respect to the population of passive galaxies in virial equilibrium
inside the clusters.
 
It is not an easy task to differentiate observationally backsplash galaxies 
from infalling galaxies, although they are possibly distinguishable through 
kinematics (e.g. \citealt{Gill:2005, Pimbblet:2006}). 
Infalling galaxies starting from larger apocenters reach larger radial velocities at their pericenters (see fig. 9 of \citealt{Mahajan:2011}).
Recently, \citet{Haggar:2020} studied the fraction of backsplash galaxies on the outskirts of clusters using hidrodynamical simulations. They found that the fraction of backsplash galaxies depends on the the dynamical state of a cluster, as dynamically relaxed clusters have a greater backsplash fraction.
\citet{Muriel:2014} explore the 
properties of galaxies in the outskirts of clusters, selecting low and high-velocity subsamples,
according to their line-of-sight velocity relative to the cluster. They 
found that late-type galaxies in the low-velocity sample are systematically older, redder, and have 
formed fewer stars than galaxies in the high-velocity sample.

To understand how environmental processes affect galaxy properties, it is necessary to
perform a good characterisation of the population of galaxies. Several authors classify galaxies around clusters using different criteria based on their position in the Projected Phase-Space 
Diagram (PPSD), which combines the projected cluster-centric distance with the line-of-sight 
velocity relative to the cluster (e.g. \citealt{Mahajan:2011, Oman:2013,Muriel:2014,Rhee:2017}). 
In \citet[hereafter dlR21]{roger}, we present the code 
\roger~(Reconstructing Orbits of Galaxies in Extreme
Regions) that relates the two-dimensional PPSD position (2D) of galaxies to their 
orbital classification (3D). The code uses three different machine learning techniques to classify galaxies 
in and around clusters according to their PPSD position. The main aim of this work is to understand
how the inherent complexity of classifying galaxies 
according to their projected position affects the conclusions we draw about the connection between their observed properties (color, star formation rate, age) and their inferred dynamical state. To carry out this study, we use a sample of galaxy clusters generated by applying the semi-analytic model of galaxy formation \sag~\citep{cora_sag_2019} to
a set of massive dark matter haloes extracted from the 
{\sc MultiDark Planck 2} cosmological simulation
 \citep{klypin_mdpl2_2016}. Firstly, we assign two orbital classifications to each galaxy, one given by their real orbit provided by the 3D phase-space and the other inferred from the PPSD through the code \roger.  
 Then, for a given galaxy class, we compare the distributions of galaxy properties (color, specific star formation rate, age) with respect to their stellar mass obtained for these two orbital classifications. Thus, we are able to evaluate the limits imposed by the 2D projections upon the conclusions that can be drawn from the observations.
We also focus on the properties of galaxies classified according to their real orbits to find out how the environment affects galaxies when they cross the virial radius of a cluster.

This article is organised as follows. 
In section \ref{sec:data}, we describe the 
sample of simulated galaxy clusters and their surroundings, the caracterisation of galaxies according to their orbits, and how we use the code \roger \ to classify galaxies in the PPSD.
The main results of the paper are presented in section \ref{sec:comparing}, where we compare the properties of galaxies belonging to different dynamical classes and 
analyse the biases introduced by the 2D orbital classification. Finally, we present our conclusions  and discussions in section \ref{sec:discussion}.

\section{Galaxy clusters and their surroundings}
\label{sec:data}

The sample of simulated galaxy clusters and of galaxies in their surroundings used throughout this work is generated by applying the semi-analytic model of galaxy
formation and evolution \sag~(acronym of Semi-Analytic Galaxies; \citealt{Cora:2018}) to the {\sc MultiDark Planck 2} cosmological simulation (\mdpl; \citealt{klypin_mdpl2_2016}). Here, we describe briefly the \mdpl~simulation, the \sag~model and the selection criteria used to construct the sample.

\subsection{The \mdpl~cosmological simulation}\label{sec:simu}

The \mdpl~cosmological simulation 
is part of the {\sc MultiDark} suite of cosmological simulations \citep{riebe_multidark_2013,klypin_mdpl2_2016}. 
This simulation counts with $3840^3$ dark matter particles in a comoving cubic box of 
$1 h^{-1} {\rm Gpc}$
of side length, resulting in a particle mass
$m_{\rm p} = 1.51 \times 10^9 h^{-1} {\rm M_{\odot}}$. The adopted cosmology corresponds to a flat $\Lambda$CDM model with $\Omega_{\rm m} = 0.307$, $h=0.678$, $n=0.96$ and $\sigma_8=0.823$, in concordance with {\it{Planck}} measurements \citep{planck_cosmology_2014,planck_cosmology_2016}.  The 
simulation tracks the dynamical evolution of dark matter particles
from the initial redshift $z=120$ using {\sc Gadget-2} \citep{springel_gadget2}. The outputs were stored into 126 snapshots between $z=17$ and $z=0$.

Dark matter haloes in \mdpl~were identified using the phase-space halo finder {\sc Rockstar} \citep{behroozi_rockstar_2013}, and the merger trees were constructed using {\sc ConsistentTrees} \citep{behroozi_trees_2013}. The halo catalogues and merger trees used in this work are publicly available at the {\sc CosmoSim}\footnote{https://www.cosmosim.org/cms/simulations/mdpl2/} and {\sc Skies \& Universes }\footnote{http://skiesanduniverses.org/Simulations/MultiDark/} databases. They are the backbone of the semi-analytic model to generate the galaxy population.

\subsection{The \sag~Model}
\label{sec:sag}

The \sag~model originates from the semi-analytic model described
in \citet{Springel+2001}. The latest version of the model is presented in \cite{Cora:2018}.
The evolution of galaxy properties is followed by assigning one galaxy to each dark matter halo detected in the underlying dark matter simulation,
and using the merger trees of haloes and subhaloes.
\sag~includes 
radiative cooling of the hot halo gas in central and satellite galaxies, quiescent star formation in the gaseous disc, 
starbursts triggered by galaxy mergers and disc
instabilities, which contributes to the formation of 
galaxy bulges \citep{lagos_sag_2008, munnozarancibia2015, gargiulo_sag_2015},
a detailed treatment of
chemical enrichment considering the contribution from 
stellar winds and different types of 
supernovae \citep{cora_sag_2006, collacchioni_sag_2018}, and feedback from supernova  and active galactic nuclei (AGN). The modelling of the former takes into account an explicit
dependence of the reaheated mass on redshift based on relations measured from 
full-physics hydrodynamical simulations; besides, gas from haloes is ejected and reincorporated later to avoid excess of star formation at high redshifts \citep{Cora:2018}.
The implementation of AGN feedback considers the growth of supermassive black holes in the centre
of galaxies \citep{lagos_sag_2008, cora_sag_2019}.

Environmental effects like tidal stripping (TS) and ram pressure stripping (RPS)
are also included in the model \citep{Cora:2018}; they are particularly relevant for the interpretation of results obtained for galaxies classified according to their real orbits (3D classification; see Section~\ref{sec:3D}).
When Galaxies become satellites, they lose their hot gas in a gradual way by the action of RPS. The gradual stripping of the hot gas is modelled assuming a spherical isothermal density profile for the gas and following 
\citet{Font+2008} and \citet{McCarthy+2008}.
When the ratio between the hot gas mass and the baryonic mass 
of a satellite galaxy decreases below $0.1$, ram pressure can strip 
gas
from the galaxy disc following the model of \citet{Tecce+2010}, where the stripping radius is calculated by using the \citet{GG:1972} 
condition.
In any case, the ram pressure exerted on the gaseous components of satellite galaxies is estimated from an analytic fitting profile that recovers the ram pressure dependence on halo mass, halocentric distance and redshift using the information
of the gas particle distribution in hydrodynamical simulations of groups and clusters
of galaxies \citep{VegaMartinez+2021}.
The mass and metals removed by RPS from both the hot gas halo and cold gas disc of the satellite are transferred
to the hot gas of the galaxy identified as the central of the 
processed satellite (i.e. the intra-cluster/group medium), which is
the central galaxy of the main host halo in most of the cases.
\sag~also considers ejection of gas from the haloes as a result of supernovae feedback, but the ejected reservoir is not affected by RPS because of 
the uncertain physical interpretation 
of this galaxy component. The rest of the galaxy components (gaseous halo and disc, stellar disc and bulge) are also affected by TS but, in general, RPS dominates.

The modelling of these physical processes involves free parameters that are calibrated to a set of observed 
relations of galaxy properties by using the 
\textit{Particle Swarm Optimisation} technique \citep{ruiz_sag_2015}. The galaxy properties considered are the stellar mass functions at redshifts $z=0$ and $z=2$, the star formation rate distribution function at $z=0.15$, the fraction of mass in cold gas as a function of stellar mass at $z=0$, and the relation between bulge mass and the mass of
the central supermassive black hole at $z=0$. The values of the set of free parameters that characterise the version of the model used here are given in table 1 of \cite{Cora:2018}, except the parameter that regulates the redshift dependence of the supernova feedback, which has been reduced to achieve a better agreement between simulated and observed values of the fraction of quenched galaxies as a function of stellar mass and halo mass (see fig. 11 in \citealt{Cora:2018} and the corresponding discussion).

It is worth noting that the calibration of the model takes into account orphan satellite galaxies, i.e., satellite galaxies whose dark matter substructure is not longer detected by the halo-finder. However, orphan galaxies are excluded from the sample considered for the analysis carried out in this work, as we did in dlR21.
The orbits of orphan satellites are tracked by an orbital evolution model in a pre-processing step before applying \sag~and, thus, they are model dependent.
Therefore, we only consider central galaxies and satellite galaxies that keep their dark matter substructure whose trajectory is well followed by the dark matter simulation.

\subsection{Sample of simulated galaxies in and around clusters}
\label{sec:sample}

The sample of clusters used in this work is the same used by dlR21. It contains 34 massive, relaxed and isolated galaxy clusters selected from the \mdpl-\sag~galaxy catalogue.
Firstly, we select all haloes at redshift $z=0$ with mass $M_{\rm 200}\ge 10^{15}h^{-1}M_{\odot}$ that have no companion haloes within $5\times R_{\rm 200}$ more massive 
than $0.1\times M_{\rm 200}$, where
$M_{\rm 200}$ is the mass contained within the region of radius $R_{\rm 200}$ that encloses 200 times the critical 
density. These criteria exclude haloes undergoing a major merger, or interacting 
with a massive companion, which may affect galaxy 
orbits in the vicinity of galaxy clusters.

We also include in our analysis those galaxies residing in the surroundings of each selected cluster. For each cluster, we consider a region delimited by a cylinder elongated along the $Z-$axis
of the simulation box which is adopted as the line-of-sight direction, without loss of generality. This region includes the cluster galaxies, galaxies in
the surroundings of the cluster, and interlopers. The latter are galaxies that 
appear in or around the cluster in projection but are unrelated to it;
they constitute a major source of contamination in the PPSD.
The dimensions of these cylinders are: a 
radius of $5\times R_{\rm 200}$, and a longitude in the $Z-$axis that extends 
as far as to include all galaxies within $|\Delta V_Z+H_0\Delta Z|\le 3\sigma$,
where $\Delta V_Z$ is the galaxy peculiar velocity in the $Z-$direction relative
to the cluster, $\Delta Z$ is the proper distance between the galaxy and the 
cluster centre in the $Z-$direction, $H_0=67.8\,{\rm km}\,{\rm s}^{-1}\,{\rm Mpc}^{-1}$
is the Hubble constant, and $\sigma$ is the one dimensional velocity dispersion
of the cluster. 
As shown by \citet{Munari:2013}, the value of $\sigma$ is sensitive to
the tracers used in its computation, i.e., dark matter particles or subhaloes. We use the
$\sigma$ values computed by dlR21. They measure the velocity dispersion of the clusters
in our sample using as tracers those subhaloes that host a central galaxy (satellite galaxy) more massive than 
$\log(M_{\star}^{\rm min}/ h^{-1}M_{\odot})=8.5$.
This threshold in stellar mass was adopted to guarantee completeness in the sample of 
galaxies \citep{Knebe2018}. Below this mass limit, observed galaxy properties are not 
followed reliably.

\subsection{Classification of galaxies in and around clusters based on their real orbits: 3D
classification} \label{sec:3D}

We classify galaxies in and around clusters according to their real orbits in 3D
using the same scheme as in dlR21. 
We consider the galaxies located within the region around each cluster of our
sample and track their positions relative to the cluster centre in the 
successive outputs of the simulation. Thus, we define five orbital classes at 
redshift $z=0$:

\begin{enumerate}
    \item Cluster galaxies (CL): 
    These galaxies involve the central galaxy of the cluster, plus all galaxies that
    have been satellites of the cluster for more than $2\,{\rm Gyr}$.
    The latter have been either inside $R_{200}$ since they have been accreted by the cluster
    or they have crossed $R_{200}$ more than twice, or just once in their way in and earlier 
    than $2\, {\rm Gyr}$ ago. Most of them are located within $R_{200}$, and 
    a few are found temporarily outside $R_{200}$ in their orbital motion.
    \item Recent infallers (RIN): 
    These galaxies are located within $R_{200}$ and have crossed this radius
    inwards only once in the last $2\,{\rm Gyr}$.
    \item Backsplash galaxies (BS): 
    These galaxies are found outside $R_{200}$ and have crossed this radius
    exactly twice in their lifetimes, the first time on their way in, and the second
    on their way out of the cluster. According to this selection criterion,
    a RIN can become a BS in the future.
    \item Infalling galaxies (IN): 
    These galaxies are in haloes that have negative radial velocity relative 
    to the cluster centre and have never been within $R_{200}$.
    \item Interlopers (ITL): 
    These galaxies have never approached the cluster centre within $R_{200}$,  
    and they are found in haloes that are receding from the cluster.
    They are not related to the cluster but, as a result of projection effects, 
    they can be  confused  with galaxies in classes i$-$iv above in the PPSD.
\end{enumerate}
Hereafter, we will refer to this classification of galaxies as 3D classification.


\subsection{Classification of Galaxies based on the Projected Phase Space Diagram Using 
\roger: 2D classification}\label{sec:2D}

In this section, we present a brief summary of the code \roger~presented in 
dlR21 and used in the present work
to carry out the orbital classification of galaxies according to their projected phase space position.

The \roger~ code is an automatic machine learning algorithm that determines the probability of each galaxy of belonging to the previously defined orbital classes using only their PPSD position.
This code was presented as an \texttt{R} 
package\footnote{\href{https://github.com/Martindelosrios/ROGER}{https://github.com/Martindelosrios/ROGER}} that can be installed and used directly from an \texttt{R} console.

To train and test \roger, dlR21 used those galaxies within 33 out of the 34 regions around the massive, relaxed and isolated galaxy clusters in our sample (Section~\ref{sec:sample}). 
With these galaxies they built two independent subsets, namely: a training-set used to train the machine learning models and a validation-set used to measure the performance of the models in each training epoch.
In turn, the galaxies belonging to the remaining galaxy cluster were used as a testing set to measure the performance of the models after the full training.
To avoid any over-fitting that may bias the present results, in this work we only use the galaxies corresponding to the  testing and validation sets originally presented in dlR21.

Although \roger~ has the possibility of using $3$ different machine learning techniques (\texttt{random forest}, \texttt{k-nearest neighbours} (\texttt{KNN}) and \texttt{support vector machines}), in the present work we use only the \texttt{KNN}  implementation as was suggested by dlR21.

To perform an analysis of the galaxy properties that resembles 
what can be done in real observations, we compute for each galaxy of our sample 
its position in the PPSD.
Phase space positions of galaxies relative to their parent cluster’s centre are computed by projecting the 3D cartesian coordinates of the galaxies in the \mdpl~box at redshift $z=0$ into the $(X, Y)$ plane. Hereafter, the projected 
distance on this plane, $R_{\rm p}$, will be referred to as the 2D distance, and it will be quoted in units of $R_{200}$ unless otherwise specified. On
the other hand, the $Z-$axis velocity relative to the cluster,
$\Delta V_{\rm los} \equiv |\Delta V_Z + H_{0}\Delta Z|$, is the line-of-sight velocity,  and 
will be quoted in units of $\sigma$. 
Then, using the previously trained code \roger~(dlR21), we compute the 
probability of each galaxy to belong to each category.

As specified in dlR21, galaxies can be classified into the five classes defined 
in Section~\ref{sec:3D} from the probabilities computed by \roger~by choosing different thresholds. 
For instance, if the $i-$th galaxy has a probability $p_{\rm cl}^{(i)}$ of being a cluster galaxy, it will be classified as such only if the probability of being a cluster galaxy is the highest probability and if it is higher than the chosen threshold, i.e., $p_{\rm cl}^{(i)} \geq T_{\rm cl}$.
In this way, we can tune up the thresholds to end up with purer, although smaller, samples, or, conversely, end up with larger but more contaminated samples.
To quantify the properties of the resulting classifications, dlR21 define the following statistics:

\begin{itemize}
    \item Sensitivity: the number of correct predictions of a given class in the resulting sample, divided by the total number of galaxies of that class in the validation-set.

    \item Precision: the number of correctly predicted galaxies of a given class in the resulting sample, divided by the total number of galaxies of that class in the resulting sample.
\end{itemize}

\begin{table}
    \centering
    \begin{tabular}{l|c|c|c|c|c}
        Class & CL & BS & RIN & IN & ITL \\
        \hline
        Real Number & $396$ & $684$ & $414$ & $1263$ & $1214$  \\
        Predicted Number & $539$ & $392$ & $286$ & $704$ & $798$\\
        Threshold & $0.4$ & $0.48$ & $0.37$ & $0.54$ & $0.15$ \\
        Precision & $0.39$ & $0.48$ & $0.55$ & $0.64$ & $0.85$ \\
        Sensitivity & $0.54$ & $0.27$ & $0.38$ & $0.35$ & $0.55$ \\
    \end{tabular}
    \caption{Sizes and thresholds values used for each orbital class and their corresponding precision and sensitivity values.}
    \label{tab:thresholds}
\end{table}

\begin{figure*}
\centering
\includegraphics[width=2.1\columnwidth]{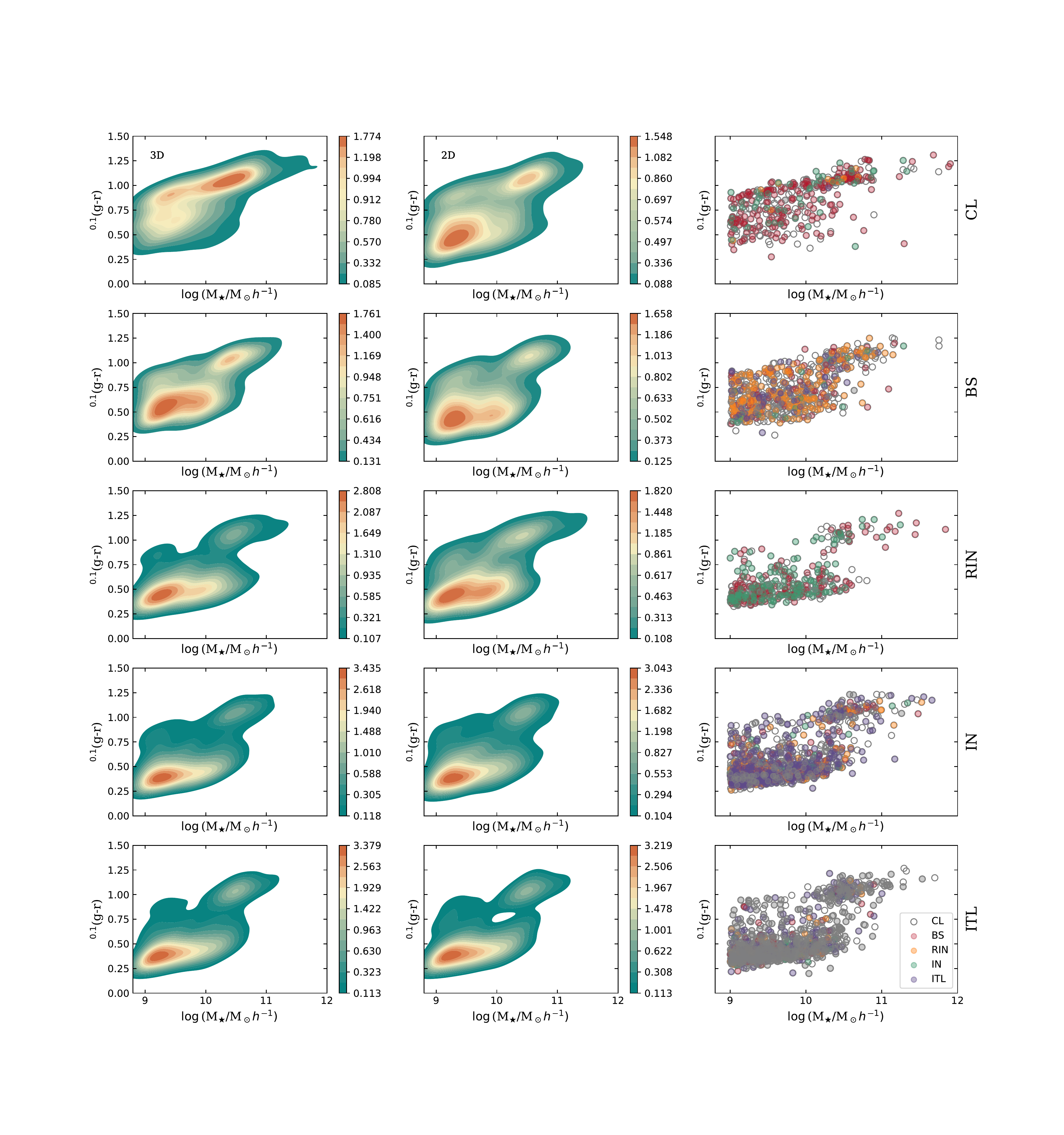}
\caption{The $\gr$ colour$-$stellar mass diagram. From \textit{top} to \textit{bottom}, the panels of each row corresponds to CL, BS, RIN, IN, and ITL galaxies, respectively. \textit{Left} panels correspond to galaxies classified according to their real orbits (3D, see Sec. \ref{sec:3D}), while \textit{central} panels correspond to galaxies classified out of their \roger~estimated probabilities (2D, see Sec. \ref{sec:2D}). In those panels, different colours indicate the galaxy number density, from the lowest (green) to the highest (orange) value in each case. 
\textit{Right} panels show in open circles the individual positions in the colour-mass diagram of the galaxies in the 3D classes. Each circle is filled with a colour corresponding to the 2D classification: CL in red, BS in orange, RIN in green, IN in violet, and ITL in grey; those circles that remain empty are galaxies that do not overpass the selection criteria and, therefore, they are not taken into account.}
\label{fig:gr}
\end{figure*}

Classifying galaxies by choosing different probability thresholds
results in samples of different degree of sensitivity and contamination. We have probed different threshold 
values and kept those that, for each class, give purer samples (high precision) without loosing statistical power 
(high sensitivity). This exploration is equivalent to moving along the curves depicted in Fig. 5 of dlR21 which 
show the sensitivity and precision of the resulting samples for different probability threshold values. 
The usefulness of such curves can be understood by considering regions in the PPSD where different real classes
coexist. For a galaxy located in one of these regions, the machine learning algorithm will predict 
similar probabilities for the coexisting classes, and hence, the galaxy will have two or more predicted 
probabilities of comparable magnitudes. If the probability thresholds are increased enough, such regions can be 
discarded, and, on the one hand, the contamination will decrease, while on the other hand, we pay the price of 
decreasing the sensitivity of the resulting samples. In the end, the trade-off between contamination and sensitivity is user dependent.

Table \ref{tab:thresholds} shows the number of galaxies belonging to each orbital class, together with the thresholds values used for each class and their 
corresponding precision and sensitivity.
As a result of this procedure, galaxies in our sample has two classifications: their (true) 3D classification based on their real orbit described in section \ref{sec:3D}, and a 2D classification, described in this section, based on the probabilities
computed by \roger~and our selected threshold values.

Although with these threshold values, miss-classified galaxies are minimised, there is still
some contamination in the final samples for each category. 
Hence, these miss-classified galaxies will introduce biases on the final properties of the 
different categories, that must be taken into account in any analysis. 
We remark that, although in the current work we focus the analysis on the results obtained using the \roger~code, these biases will be present in any study of galaxy properties that tries to make a dynamical classification from the PPSD positions since in real observations the full 3D orbits can not be determined.
The analysis of the biases  introduced by the 2D 
classifications in the statistical properties that characterise
each category is the main contribution of this work and will be the main topic of the 
following sections.

\section{Comparing properties of galaxies: 3D vs 2D classifications}\label{sec:comparing}
\begin{figure*}
\centering
\includegraphics[width=2.1\columnwidth]{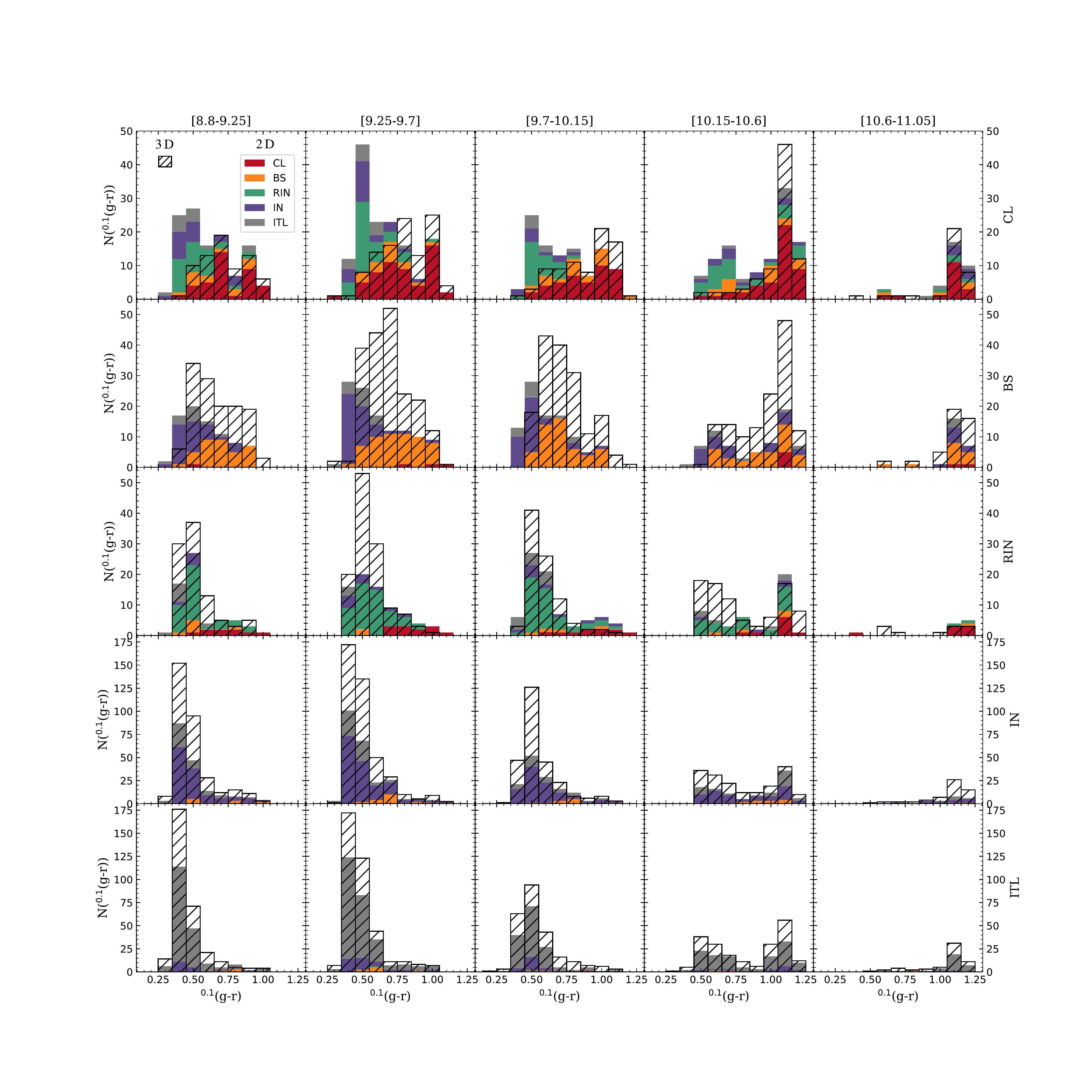}
\caption{Distribution of $\gr$ colour in five bins of stellar mass. 
Dashed histograms correspond to galaxies classified according to their real orbit, 
while shaded histograms  correspond to the 2D classification.
Galaxies classified by \roger~ are shown as shaded coloured histograms. 
Each colour corresponds to a different 3D class as specified in the inset box.}
\label{fig:barplot}
\end{figure*}

 In this section, we focus on the dependence on stellar mass of the $\gr$ colour, the specific star formation rate (sSFR), and the stellar age of galaxies, analysing the trends obtained for different orbital classes, and compare the results obtained from the 3D and 2D classifications.
The results for the 3D classification are interpreted according to the environmental processes included in \sag~(Section~\ref{sec:sag}).


\subsection{Colour}\label{sec:gr}

\begin{figure*}
\includegraphics[width=2\columnwidth]{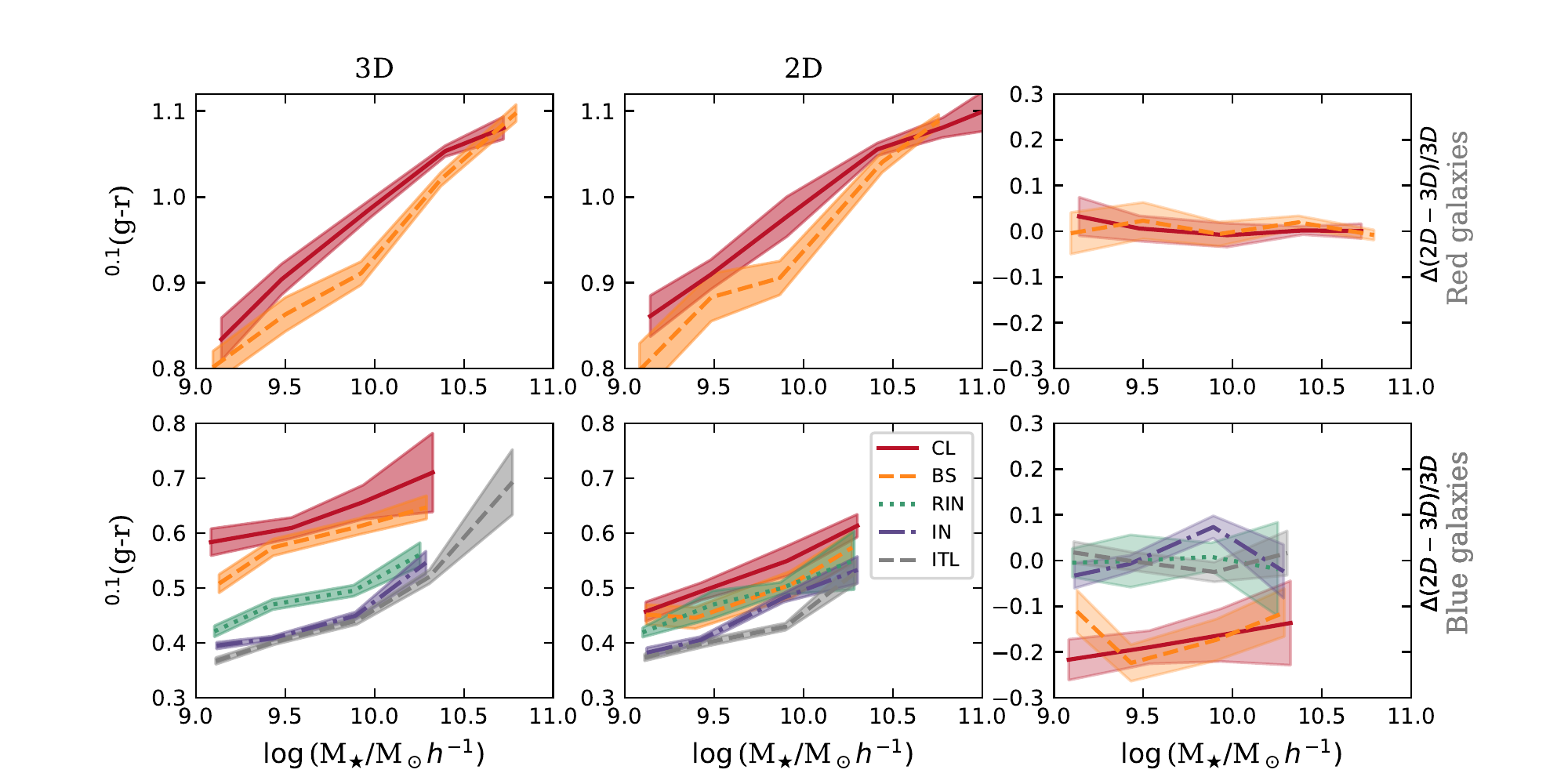}
\caption{
Median $\gr$ colour as a function of
stellar mass. Shaded areas depict errors obtained by the bootstrap resampling technique. \textit{Top} panels correspond to red galaxies and \textit{bottom} panels show blue galaxies.
\textit{Left} panels correspond to galaxies selected by their real orbits (3D), and \textit{central} panels show the results for
galaxies classified according to the \roger~ code (2D). 
\textit{Right} panels show the colour relative difference
between the 3D and 2D orbital types for each stellar mass bin, with shaded areas showing the propagation of the errors from the 3D and 2D classes.}
\label{fig:median_gr}
\end{figure*}

Fig. \ref{fig:gr} shows the colour-stellar mass diagram for each galaxy class. 
We have adopted a band shift to a redshift $z=0.1$ for the colour to approximate the mean redshift of the main galaxy sample of SDSS \citep{York:2000}.
Left panels correspond to 3D classes, while central panels to 2D classes, classified out of their \roger~ estimated probabilities. Different colours indicate the galaxy number density, from the 
lowest non-zero value (green) to the highest value (orange). On the other hand, right panels show in open circles the individual positions in the colour-mass diagram of the galaxies in the 3D category, where each circle is filled with a colour corresponding to the 2D classification. From top to bottom, we show CL, BS, RIN, IN and ITL galaxies, respectively; we observe a gradual increase of the blue population for both, 3D and 2D classes. For 3D classes, we see that the number of BS galaxies lying on the red sequence is lower than the corresponding number of CL galaxies, on one hand, and higher than the number of red RIN galaxies, on the other.
If we consider the 2D classes, there is no clear difference 
in the galaxy distribution on the colour-stellar mass plane between BS and RIN.  
Right panels of Fig. \ref{fig:gr} show how the colour-mass diagrams of the 2D 
classification are built up in terms of galaxies in the five 3D (real) classes.
As can be seen from this figure, the most common miss-classifications occur between
CL and RIN galaxies, and between BS and IN galaxies. 
Interlopers contaminate almost exclusively the IN class. It is worth 
mentioning that this behaviour was already pointed out by dlR21.
These miss-classifications have immediate and important consequences for the 
2D classes: CL galaxies appear to have a relevant blue cloud that is not real but a
result of the contamination by bluer RIN galaxies; BS galaxies have a weaker red
sequence due to contamination from bluer IN galaxies.
 
A more detailed analysis on how the 2D classes are made up out of the five 3D classes as a function of the stellar mass is shown in Fig. \ref{fig:barplot}. In this figure, we split our samples of galaxies into five bins of stellar mass: 
$\log(M_{\star}/{\rm M}_{\odot}h^{-1})=[8.80-9.25]$, $[9.25-9.70]$, $[9.70-10.15]$, $[10.15-10.60]$, and $[10.60-11.05]$, and show the corresponding $\gr$ colour distributions.
We observe a fictitious excess of blue galaxies in the 2D classes CL and BS over the
whole mass range. In the case of CL galaxies, this blue excess is made up mostly by misclassified RIN galaxies. For BS galaxies, the main source of the excess are misclassified IN galaxies.
In addition, we see that a large number of 3D BS galaxies are lost. They are either, classified as other class, or left out of the 2D classification as their computed probabilities of being BS do not exceed the chosen threshold.
Fig. \ref{fig:barplot} warns about straightforward interpretations of observations
when galaxies are classified according to their position in the PPSD. The blue population 
of galaxies classified as CL or BS will have a high degree of contamination thus affecting the  conclusions that can be drawn. 
Nonetheless, we note that the red population of CL and BS galaxies are well recovered by 
\roger. On the other hand, RIN, IN and ITL galaxies have a good agreement between the 3D 
and 2D types. It is worth emphasising that, although we are analysing the results obtained using the \roger~code, any code that classifies galaxies according to their PPSD position is likely to have similar biases.

In the light of Fig. \ref{fig:barplot}, it is strongly recommended a separate analysis of the red sequence and the blue cloud whenever possible while studying galaxy properties in and around clusters.  To separate our sample in red and blue populations, we consider a threshold linear in the logarithm of stellar mass,  $\gr=0.125\times\log(M_\star/M_\odot h^{-1})-0.45$, adopted by Salerno et al. (2021, in preparation), based on the $\gr$ colour–stellar mass diagram for galaxies in the \sag~model at $z=0$. In Fig. \ref{fig:median_gr} we show the median value of the colour as a function of the stellar mass.
 Median colours are computed by splitting galaxies into the same stellar mass bins used in Fig.~\ref{fig:barplot}. We restrict to those bins containing at least ten galaxies. Shaded areas were computed using the bootstrap re-sampling technique. Given the low number of red galaxies in 2D 
types for the RIN, IN and ITL classes, we only analyse the properties of
the red population of galaxies for CL and BS classes, hereafter.

\begin{figure*}
\includegraphics[width=2\columnwidth]{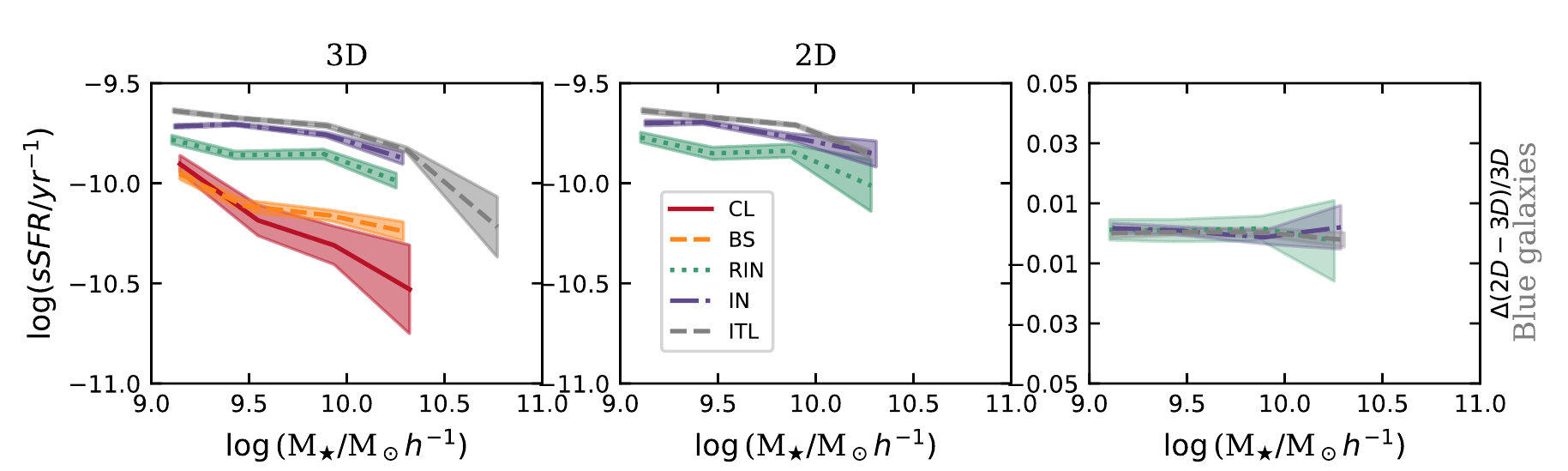}
\caption{
Median sSFR as a function of stellar mass
for blue galaxies.} 
\label{fig:ssfr}
\end{figure*}

From Fig.~\ref{fig:median_gr},
we observe an increase of the median colour with increasing stellar mass for both 
the red and blue populations, when considering either the 3D or the 2D orbital classes (left and central panels, respectively). More specifically, for a fixed stellar mass, CL 
galaxies are systematically the reddest. In particular, for the red population of galaxies, there are no significant differences in colour between real and projected orbital types, as we show in the right panels of Fig.~\ref{fig:median_gr}. We can see that the largest bin-to-bin differences are of the order of $\sim 3$ per cent and $\sim 2$ per cent, for CL and BS galaxies, respectively.
For the blue population of the 3D types,
there is a clear difference in colours among the galaxy types for a fixed 
stellar mass: CL galaxies are the reddest, followed by BS and RIN galaxies, in that order. IN and ITL galaxies are the bluest with small differences between them. 
However, we observe an important difference in colours between the 3D and 2D blue populations of classes CL and BS, being  up to $\sim 20$ per cent for both, CL and BS galaxies. This is a consequence of the contamination by a the fictitious blue population discussed above (see Fig. \ref{fig:barplot}).
This discrepancy is not seen for the other classes, where the largest bin-to-bin differences are $\sim 2$ per cent for RIN and ITL galaxies, and $\sim 7$ per cent for IN galaxies.

The results shown in Fig.~\ref{fig:median_gr} strongly suggest that when studying CL and BS galaxies in observations, only their red populations can be reliably studied.
On the other hand, studies of the blue population are reliable for RIN, IN, and ITL classes. 
Our samples do not allow for a reliable statistics of the red population of these latter classes.

\subsection{Specific star formation rate and stellar age}\label{sec:properties}

\begin{figure*}
\includegraphics[width=2\columnwidth]{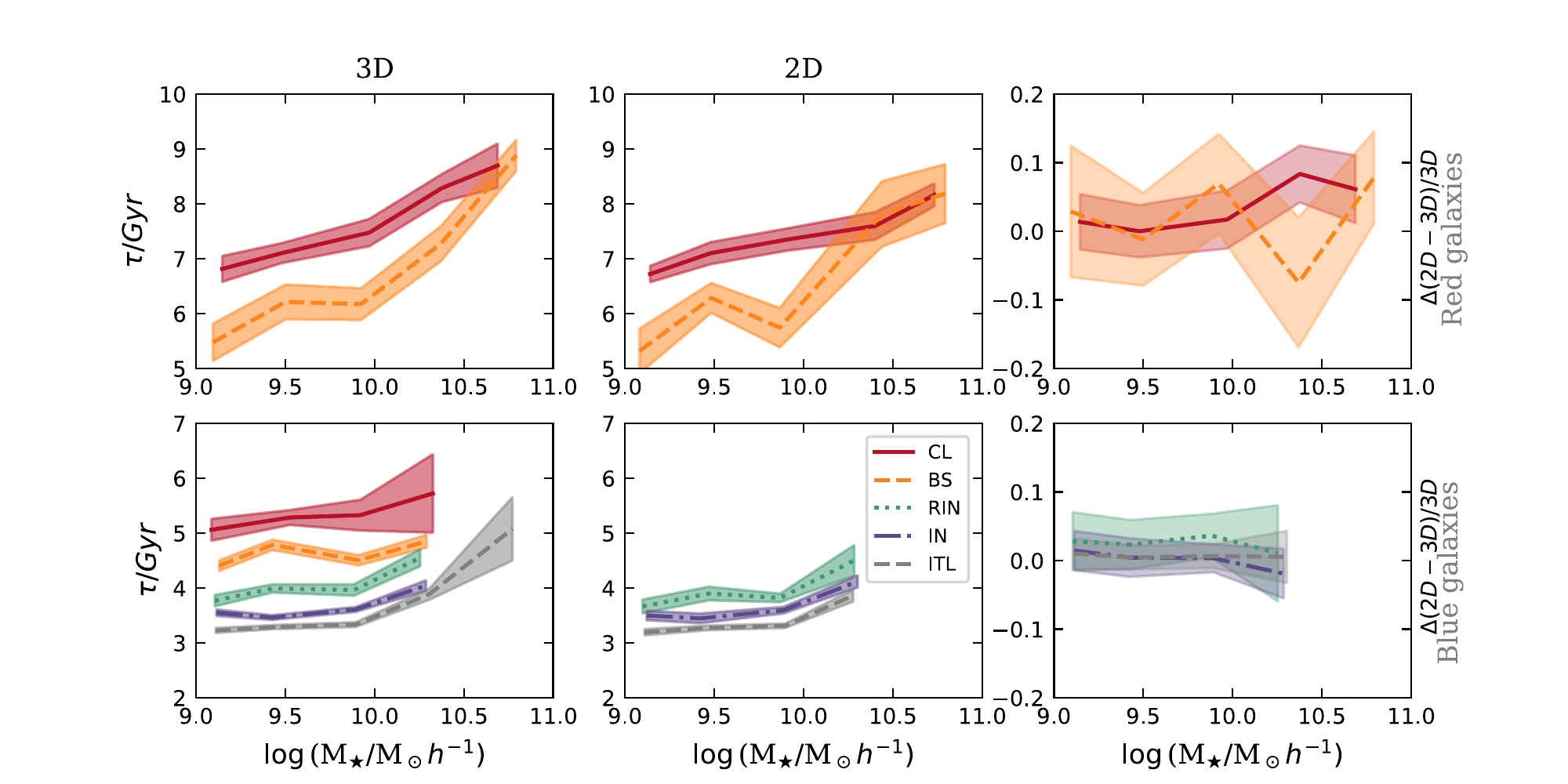}
\caption{Median of stellar age $\tau$ as a function of the stellar mass. }
\label{fig:age}
\end{figure*}

Galaxy colours are a natural consequence of the evolution of the galaxy star formation rate, which also determines its average stellar age. Thus, we now analyse the sSFR and age of the red an blue galaxy populations, separately. As in Fig. \ref{fig:median_gr}, we split each
sample in five stellar mass bins. In the light of the results in Sect. \ref{sec:gr}, the CL and BS galaxies have not been taken into account to analyse the blue populations of galaxies.

In Fig. \ref{fig:ssfr}, we compare the median values of the
sSFR as a function of the stellar mass, for the blue population of galaxies in all five classes. We do not present the median values of sSFR for the red populations of galaxies
because most of them have already ceased their star formation in their evolution. As those red galaxies have not formed new stars during the last time-step considered to calculate their sSFR, their null values of star formation rate can not be included in the current analysis. Moreover, the fraction of red galaxies featuring star formation is not large enough to do a robust comparison using them, as each stellar mass bin is filled with less than 10 galaxies.

As expected, the sSFR decreases with increasing stellar mass (e.g. \citealt{Spindler:2018, Belfiore:2018, Coenda:2019}); this behaviour is observed independently of the galaxy class. For 3D classes, CL galaxies are the most passive systems, followed by BS types. RIN types are in a intermediate zone between BS and IN types. 
This explains the progressive redder colours displayed by these 3D classes, with IN being the bluest galaxies and CL the reddest ones.

If we focus only on the RIN, IN and ITL types, as suggested in the previous section, we do not observe a significant difference between the 3D and 2D classes, 
being the largest bin-to-bin differences of $\sim 2$ per cent for RIN and IN classes of galaxies, and $\sim 0.05$ per cent for ITL galaxies.

Finally, in Fig.~\ref{fig:age}, we analyse the median of stellar age $\tau$ as a function of the stellar mass. For the red sub-sample, CL galaxies are systematically the 
oldest. Only the highest stellar mass bin shows that BS and CL galaxies have similar stellar ages. The 2D trends  are consistent with the 3D predictions. The bin-to-bin comparison shows that the differences between 2D and 3D are lower than $\sim 8$
per cent and almost consistent with zero considering the large errors.
For the blue sub-sample of 3D types, CL galaxies are clearly the oldest ones
followed by BS, RIN, IN and ITL galaxies, in that order. It is interesting to note that blue CL and BS galaxies differ significantly in $\tau$ in contrast to their similarity regarding colour or sSFR.
For 2D types, we observe the same behaviour for RIN, IN and ITL types.
The largest difference between 3D and 2D classes occurs for RIN galaxies, with a maximum value of $\sim 3$ per cent, while the largest difference for IN and ITL galaxies is $\sim 2$ per cent and $\sim 1$ per cent, respectively.


\section{Discussion and Conclusions}\label{sec:discussion}

In this paper we perform a comparison of properties of galaxies around massive and isolated clusters 
galaxies at $z=0$, using a sample of galaxies generated from the combination of the semi-analytic 
model of galaxy formation (\sag) and the MultiDark Planck 2 simulation (MDPL2). We track the orbit of the 
galaxies in our sample to define five galaxy classes: cluster galaxies (CL), galaxies that have recently 
fallen into a cluster (RIN), backsplash galaxies (BS), infalling galaxies (IN), and interlopers (ITL). 

Since only galaxy positions in PPSD are accessible, different methods have been proposed to link these positions with orbits (e.g. the inversion of the Jeans equation, \citealt{Binney:1982}; \citealt{Natarajan:1996}; \citealt{Biviano:2004}). We use the probabilities computed by \roger, to make the best use of the PPSD information and classify our sample of galaxies into the five classes mentioned above.
We choose different thresholds values for each class to 
minimise the overlapping regions in the projected phase-space and hence end with more pure samples.
We analyse a number of galaxy properties: $\gr$ colour, the sSFR and the stellar age, as a function of the stellar mass, 
and consider separately red and blue sub-samples of galaxies 
for both the 3D classes (classified according to their real orbits) and the 2D classes (classified with \roger).

Environmental effects inherent to galaxy clusters are clearly evident when focusing on the properties (colour, specific star formation rate, age) of 3D classes.
Our results show that,
for the red population and at fixed stellar mass,
galaxies in clusters are redder and older than BS galaxies. For the blue population, CL galaxies are the reddest, the oldest and have the lowest specific star formation rate. When a galaxy dives into a cluster for the first time, environmental quenching onsets and the galaxy's stellar population gets older and redder in comparison to galaxies external to the cluster. 
The action of environmental effects is even stronger for BS galaxies as they have travelled longer inside the cluster. However, they  are not yet as red, old and with the star formation as quenched as the cluster galaxies that have experienced the environmental effects for a longer period, i.e., CL galaxies. Infalling and interlopers galaxies are quite similar; these are forming stars, are blue and have young stellar populations.

We interpret the trends depicted by galaxies classified according to their real orbits taking into account the modelling of environmental effects implemented in \sag~ (see Section~\ref{sec:sag}).
Galaxies defined as satellites by the halo finder are affected by gradual removal of hot halo gas and cold gas disc by RPS. While most of CL galaxies are within the virial radius of the cluster, there are galaxies identified as satellites by the halo finder that reside at a further clustercentric distance. Hence, not only CL galaxies are affected by such processes but BS, RIN and IN galaxies could also feel the action of RPS when approaching the cluster. Therefore, galaxies that have dived for the first time into the cluster in the last 2 Gyr (RIN) already have signatures of the environmental quenching in their star formation rate (Fig.~\ref{fig:ssfr}) and are redder than IN galaxies (Fig.~\ref{fig:median_gr}).
However, in the outskirts of clusters, the values of ram pressure are low, of the order of $\sim 10^{-12}\,h^{2}\,{\rm dyn}\,{\rm cm}^{-2}$ (see fig. 1 of \citealt{VegaMartinez+2021}).

When a satellite galaxy crosses the virial radius of a host halo and approaches the cluster centre, it is affected by increasing values of ram pressure, not only because ram pressure is higher in the inner regions of the cluster but also because of the build-up of the intra-cluster medium density over time as the satellite's orbit evolves.
At $z \sim 2$, most of satellite galaxies are already experiencing medium-level ram pressure (i.e., $\sim 10^{-11}\,h^{-2}\,{\rm dyn}\,{\rm cm}^{-2}$) and, at
$z\lesssim 0.5$, those satellites located  within $0.5 \, R_{\rm 200}$ feel strong-level ram pressure, i.e.,  $\gtrsim 10^{-10}\,h^{2}\,{\rm dyn}\,{\rm cm}^{-2}$ (\citealt{Tecce+2010, VegaMartinez+2021}).
The most important effect of RPS is the gradual removal of the hot gas component of satellite galaxies. The mass stripped by ram pressure is larger
for less massive satellites, being the
the cumulative stripped hot gas fractions of the order of $\sim 70$ per cent for low-mas satellites ($M_{\star}\sim 3\times 10^9 \, M_{\odot}$) and $\sim 40$ per cent for high-mas ones ($M_{\star}\sim 10^{11}\, M_{\odot}$; see fig. 14 of \citealt{Cora:2018}).
The lost of cold gas by RPS is an order of magnitude lower because the cold gas disc is shielded by the hot gas halo during most of the galaxy lifetime.
Hence, the final effect of RPS is the reduction of the gas cooling rate which derives in the quenching of the star formation rate, which in turn is reflected in the colours and average ages of satellites.

It is clear then that the sSFR and colour is more affected the longer a galaxy is in a cluster, giving rise to the progressive 
 quenching and reddening manifested by ITL, IN, RIN, BS and CL galaxies. 
A single excursion in and out of a cluster (BS) quenches star formation in a significant way. These results are in agreement with \citet{Lotz:2019}.
Thus, with only one passage close to the centre of the cluster, galaxy colour is strongly affected, as suggested by the comparison between BS and RIN. 

By focusing on the 2D classes, we see an important contamination over the projected populations: 
there are miss-classifications between CL and RIN galaxies, and BS and IN galaxies. Interlopers contaminate almost exclusively the IN class. These miss-classifications have significant consequences for the 2D classes: CL galaxies appear to have an exceptional blue cloud that is not real but a 
result of the contamination by bluer RIN galaxies; BS galaxies have a weaker red sequence due to contamination from bluer IN galaxies. This degree of contamination 
considerably affects the conclusions that can be deduced.
From our results, we consider that it is necessary to separate the red and blue galaxy populations to perform a more realistic analysis of observational data. Reliable results can be obtained for red CL and BS galaxies, and for blue RIN and IN galaxies. Interlopers are not affected significantly by contamination. We recall that the red population of RIN, IN and ITL galaxies has not been analysed due to the low number of galaxies, however, these classes should not be discarded in future studies.

From the observational point of view, \citet{Muriel:2014} 
explore several properties of early and late-types galaxies, defined in terms of 
their concentration index, in the outskirts of clusters. They select galaxies 
with projected distances in the range $1<R_{\rm p}/R_{\rm vir}<2$ from a sample of X-ray selected clusters (\citealt{Coenda:2009, Muriel:2014}). They distinguish between low velocity ($\Delta V_{\rm los}/\sigma<0.5$), and high velocity ($\Delta V_{\rm los}/\sigma>1$) galaxies. 
They find evidence that, for a fixed stellar mass, late-type low-velocity galaxies have formed less stars in the last 3 Gyr, are redder and older, than their 
high-velocity counterparts. In terms of their PPSD position, we can qualitatively 
associate high-velocity galaxies with the IN class, and the low-velocity galaxies with
the BS class.
Since each colour branch is also associated to a different galaxy morphology, i.e., late-type galaxies are associated to the blue cloud, whereas early-types are located in the red sequence in the colour-magnitude diagram (e.g. \citealt{Conselice:2006}),
thus, the results of \citet{Muriel:2014} are 
consistent
with the results presented in this paper.

There are some works in the literature that compare the properties of galaxies classified according to their position in the projected phase-diagram (e.g. \citealt{Biviano:2002, Oman:2013, Muzzin:2014, Lotz:2019, Lotz:2021}) but they are a few. However, there is effort to achieve a good separation of the galaxies in the phase-space diagram (e.g. \citealt{Mahajan:2011, Rhee:2017, Jaffe:2018}).  Based on the results shown here, we will apply the \roger~ code to a real sample of galaxy clusters in a forthcoming paper to shed light on the properties of galaxies in the surroundings of galaxy clusters.

\section*{Acknowledgements}
The authors thank the referee, Andrea Biviano, for his comments and suggestions that improved
the paper.
This paper has been partially supported with grants from \textit{Consejo Nacional de 
Investigaciones Cient\'ificas y T\'ecnicas} (CONICET, PIP 11220130100365CO) Argentina, the \textit{Agencia Nacional de Promoci\'on Cient\'ifica y Tecnol\'ogica} (PICT 2016-1975), Argentina, and \textit{Secretar\'ia de Ciencia y Tecnolog\'ia, Universidad Nacional de C\'ordoba}, Argentina.
MdlR acknowledges financial support from the Comunidad Aut\'onoma de Madrid through the grant
SI2/PBG/2020-00005 and FAPESP through the process 2019/08852-2. 
SAC acknowledges funding from {\it Consejo Nacional de Investigaciones Cient\'{\i}ficas y T\'ecnicas} (CONICET, PIP-0387), {\it 
Agencia Nacional de Promoci\'on de la Investigaci\'on, el Desarrollo Tecnol\'ogico y la Innovaci\'on} (Agencia I+D+i, PICT-2018-03743), and {\it Universidad Nacional de La Plata} (G11-150), Argentina.
CVM acknowledges financial support from the Max Planck Society through a Partner Group grant. 

The \textsc{CosmoSim} database used in this paper is a service by the
Leibniz-Institute for Astrophysics Potsdam (AIP). The authors gratefully
acknowledge the Gauss Centre for Supercomputing e.V. (www.gauss-centre.eu) and
the Partnership for Advanced Supercomputing in Europe (PRACE, www.prace-ri.eu)
for funding the \textsc{MultiDark} simulation project by providing computing
time on the GCS Supercomputer SuperMUC at Leibniz Supercomputing Centre (LRZ,
www.lrz.de).

\section*{Data availability}

\roger~data underlying this article are available in \textsc{Github} at \url{https://github.com/Martindelosrios/ROGER} and in \textsc{zenodo} at  
\url{https://zenodo.org/badge/latestdoi/224241400}. The raw data of the semi-analytic model of galaxy 
formation \sag~will be shared on reasonable request to the corresponding author.




\bibliographystyle{mnras}


\bsp	
\label{lastpage}
\end{document}